\documentclass[%
reprint,
superscriptaddress,
 amsmath,amssymb,
 aps,
 prapplied,
]{revtex4-2}
\usepackage{graphicx}
\usepackage{dcolumn}
\usepackage{bm}
\usepackage[dvipsnames]{xcolor}
\usepackage{comment}

\begin{document}

\preprint{APS/123-QED}

\title{Trapping photo-carriers in undoped gate-defined quantum dots using an on-chip microwave resonator for charge readout}

\author{Pierre Lefloïc}
\email{Pierre.Lefloic@USherbrooke.ca}
\affiliation{Département de Physique, Université de Sherbrooke, Sherbrooke, Québec J1K 2R1, Canada}
\author{Zhiren Wang}
\affiliation{Département de Physique, Université de Sherbrooke, Sherbrooke, Québec J1K 2R1, Canada}
\author{Yasmine Faraj}
\affiliation{Département de Physique, Université de Sherbrooke, Sherbrooke, Québec J1K 2R1, Canada}
\author{Harshana Runjeet}
\affiliation{Département de Physique, Université de Sherbrooke, Sherbrooke, Québec J1K 2R1, Canada}
\author{Michel Pioro-Ladrière}
\affiliation{Département de Physique, Université de Sherbrooke, Sherbrooke, Québec J1K 2R1, Canada}
\author{Alicia Kam}
\affiliation{National Research Council Canada, 1200 Montreal Road, Ottawa, Ontario, K1A 0R6, Canada}
\author{Louis Gaudreau}
\affiliation{National Research Council Canada, 1200 Montreal Road, Ottawa, Ontario, K1A 0R6, Canada}
\author{Mathieu Juan}
\email{Mathieu.Juan@USherbrooke.ca}
\affiliation{Département de Physique, Université de Sherbrooke, Sherbrooke, Québec J1K 2R1, Canada}


\date{\today}

\begin{abstract}
Scaling up gate-defined quantum dot systems is hindered by the rapid growth in the number of control gates. To tackle this challenge, we propose a different approach where the quantum dots are created from optically generated charges trapped beneath accumulation gates. By shining an above-the-gap laser light onto an undoped GaAs substrate, we demonstrate that it is possible to create and separate electron-hole pairs to form quantum dots. For the charge readout, we use a superconducting coplanar waveguide resonator capacitively coupled to the quantum dots. With these elements, we demonstrate a working many-charge double quantum dot device comprised of only two plunger gates, one tunnel coupling gate and one sensing gate. With our approach, we show that populating quantum dots does not require engineered reservoirs, source/drain bias, ohmic contacts, or in-situ doping. As a result, the number of gates can be kept minimal and the fabrication process can be simplified. Moreover, this approach can be applied to a wide range of direct-bandgap semiconductor quantum dot systems. This hybrid device demonstration is a first step towards a different scalability approach for quantum dot arrays. 

\end{abstract}

\maketitle


\section{\label{sec:intro}Introduction}

Scaling quantum processing units to large numbers of qubits remains one of the central challenges in the pursuit of practical quantum computation~\cite{veldhorst2017, dumoulinstuyck2026}. While impressive progress has been achieved across several physical platforms, realizing systems that combine high-fidelity control, long coherence times, and scalability remains an outstanding goal. Semiconductor spin qubits are widely regarded as a promising approach due to their compatibility with the mature semiconductor fabrication ecosystem and the possibility of leveraging existing industrial manufacturing techniques~\cite{chatterjee2021, stano2022}.

The concept of encoding quantum information in the spin of an electron confined in a semiconductor quantum dot was first proposed by Loss and DiVincenzo~\cite{Loss1998}. Subsequent experimental demonstrations of spin qubits in gate-defined GaAs quantum dots established the viability of this approach~\cite{Petta2005,Koppens2006}. However, these early devices were limited by strong hyperfine interactions with the dense nuclear spin environment of the host material, which lead to rapid spin dephasing~\cite{Petta2005,Bluhm2011}. This limitation motivated the transition toward silicon-based platforms, where the dominant isotope ($^{28}$Si) has zero nuclear spin. Advances in isotopic purification have enabled silicon devices with dramatically reduced magnetic noise~\cite{tyryshkin2012}. Despite the novel challenges associated with silicon, spin qubits have since been realized in multiple silicon architectures, including metal–oxide–semiconductor (MOS) quantum dots~\cite{veldhorst2015, veldhorst2017} and Si/SiGe heterostructures~\cite{Zajac2018,Watson2018, Mills2022, Mills2022a}. These systems have demonstrated steadily increasing coherence times~\cite{stano2022, steinacker2025} with single- and two-qubit gate operation fidelities now exceeding 99\% for both operation types~\cite{burkard2023, Yoneda2018, Noiri2022, Mills2022a, steinacker2025}.

Nevertheless, the size of controllable qubit arrays remains relatively modest. Demonstrations have so far reached 2D arrays of four coupled qubits~\cite{hendrickx2021}, ten qubits~\cite{wang2024} and sixteen qubits~\cite{borsoi2024}, and linear arrays of up to twelve qubits~\cite{george2024}. Scaling these architectures to larger systems is complicated by the large number of electrostatic gates required to define, tune, and couple individual quantum dots, leading to increased wiring complexity and control overhead. One potential strategy to alleviate these scaling constraints is to interface stationary spin qubits with flying photonic qubits, enabling long-range coupling and quantum networking~\cite{vrijen2001,wehner2018,laucht2021}. Motivated by this prospect of spin–photon interfaces, considerable effort has been devoted to optically active defect centers in semiconductors, such as nitrogen–vacancy centers in diamond~\cite{dutt2007,togan2010,pezzagna2021} and T-centers in silicon~\cite{bergeron2020, higginbottom2022, song2026}. These systems combine long-lived spin states with optical addressability and coherent photon emission. However, their scalability remains limited by fabrication challenges, particularly the deterministic placement of defects~\cite{pezzagna2021} and the variability of their optical properties~\cite{bergeron2020}. An appealing alternative is to combine the precise electrostatic tunability of gate-defined quantum dots with the optical functionality typically associated with defect-based systems. Previous work has demonstrated that the angular momentum of an absorbed photon can be transferred to a photogenerated electron, enabling the polarization state of a photon to be mapped onto an electron spin across the semiconductor bandgap~\cite{pioda2011, fujita2013, morimoto2014, gaudreau2017, fujita2019}. Such mechanisms open the possibility of realizing spin–photon interfaces within gate-defined semiconductor nanostructures.

In this work, we explore the minimal device architecture required to realize a functional double quantum dot by leveraging photogenerated charge carriers. Guided by this minimalist approach, we fabricate a gate-defined quantum dot device that operates without ohmic contacts and is formed in an undoped semi-insulating GaAs heterostructure where charge carriers are generated entirely through optical excitation. This work provides an initial demonstration of the proposed concept, focusing on charge trapping as a prerequisite for spin-qubit operation. Nonetheless, this architecture could provide a simplified platform for investigating optically induced charge and spin dynamics in gate-defined quantum dots and represents a step toward hybrid spin–photon quantum devices.

\section{\label{sec:device}Device design and description}

\begin{figure}[t]
    \centering
    \includegraphics[width=246pt]{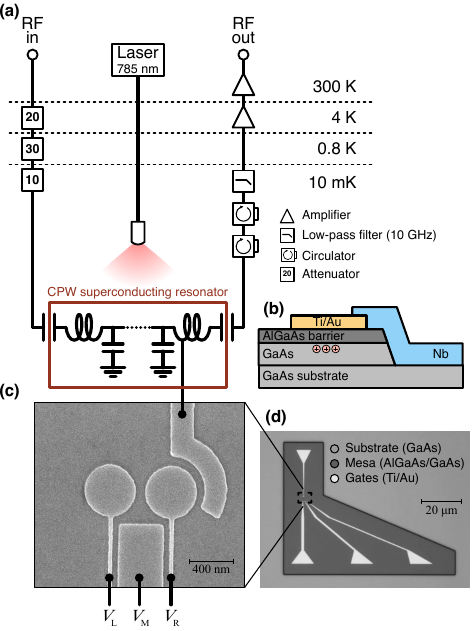}
    \caption{Microwave circuit, micrographs, and mesa schematic. (a) Microwave circuit used to measure the $S_{21}$ parameter of the superconducting coplanar waveguide (CPW) resonator, together with its lumped-element circuit model. A total of 60~dB of attenuation is applied to the microwave input line. (b) Schematic cross section of the mesa structure. The niobium film forming the resonator lies on the substrate, while only the gate leads extend along the mesa sidewall to connect to the Ti/Au gates. The quantum dots form at the AlGaAs/GaAs interface beneath the gates. (c) Scanning electron micrograph of the double-quantum-dot gate geometry. The two plunger gates are separated by a central barrier gate. A fourth gate, capacitively coupled to the right quantum dot, connects to the superconducting resonator. (d) Optical micrograph of the mesa structure, acquired before deposition of the niobium leads.}

    \label{fig:fig1}
\end{figure}

The device used in this study is a double quantum dot (DQD) formed in undoped GaAs beneath an Al$_{0.33}$Ga$_{0.67}$As barrier. The heterostructure was grown by molecular-beam epitaxy (MBE). The DQD is capacitively coupled to a superconducting coplanar waveguide resonator used for dispersive charge sensing~\cite{petersson2010, frey2012, petersson2012, colless2013, viennot2016}. Figure~\ref{fig:fig1}(c) shows a scanning electron micrograph of the electrostatic gate layout. The gate electrodes are defined using electron-beam lithography followed by a lift-off process of a Ti/Au bilayer with thicknesses of (3/15)~nm. The two plunger gates are used to electrostatically define the quantum dots, while the central barrier gate controls the interdot tunnel coupling and simultaneously depletes the charge accumulation beneath the plunger leads. The impact of this design was investigated using finite-element simulations and the results are presented and discussed in a later section (section~\ref{sec:schottky}).

In contrast to conventional gate-defined quantum dot devices, this architecture operates without ohmic contacts and relies entirely on optically generated carriers. A sickle-shaped gate is capacitively coupled to the right quantum dot and connected to the superconducting resonator, enabling dispersive readout of the dot occupation~\cite{petersson2010, frey2012, petersson2012, colless2013, viennot2016}. Superconducting coplanar waveguide (CPW) resonators have been established as reliable and non-invasive charge sensors~\cite{viennot2016}. The readout resonator is fabricated via a lift-off process of a 20~nm-thick niobium film, yielding a nominal resonance frequency of 6~GHz and a characteristic impedance of 50~$\Omega$. Microwave transmission measurements are performed by probing the scattering parameter $S_{21}$ through two capacitively coupled ports. The resonator, modeled as a lumped-element circuit, is illustrated in Fig.~\ref{fig:fig1}(a). In this context, the quantum capacitance of the double quantum dot, arising from its discrete energy spectrum, acts as an additional capacitive load on the resonator, resulting in a dispersive shift of its resonance frequency~\cite{duty2005, gonzalez-zalba2015, mizuta2017}. For sufficiently strong coupling, this interaction can be described within the framework of circuit quantum electrodynamics (cQED), where the quantum states of the resonator and the DQD hybridize~\cite{petersson2012}. The device studied here operates below this strong-coupling regime, where the interaction manifests primarily as a dispersive shift of the resonator frequency.

While the gates control primarily the lateral confinement, the vertical confinement of the quantum dots is achieved from the combined effect of the 50~nm-thick AlGaAs barrier and the electrostatic attraction induced by the metallic gates. This architecture operates similarly to an accumulation-mode device, but the absence of ohmic contacts prevents the formation of a two-dimensional gas right away. Instead, light is used to photo-dope the heterostructure, providing the charge carriers. These accumulate in the GaAs layer beneath the barrier and are confined at the interface, resulting in the formation of a quasi-two-dimensional charge gas in a triangular potential well at the AlGaAs/GaAs interface. This confinement is further reinforced by the presence of a Schottky barrier at the metal–semiconductor interface, as discussed in Sec.~\ref{sec:schottky}. Owing to the absence of intentional doping and the use of optically generated carriers, the device architecture should, in principle, support both electron and hole operation~\cite{plentz1997, glasberg2001}. However, only hole-based quantum dots are investigated in the present work.

To populate the quantum dots, electron–hole pairs are generated optically across the bandgap within the semiconductor. In principle, this process also creates carriers beneath the resonator, where they can become trapped at the heterointerface due to the Schottky potential. Such unintended charge accumulation introduces additional losses and degrades the resonator performance. To mitigate this effect, the resonator is fabricated directly on the GaAs substrate~\cite{sun2018}. This is achieved by selectively etching away the surrounding heterostructure, leaving the epitaxial layers only in the double quantum dot region and thereby forming a mesa structure, as shown in Fig.~\ref{fig:fig1}(d). Figure~\ref{fig:fig1}(b) shows a schematic cross section of the heterostructure. The niobium film forming the resonator extends along the mesa sidewall to make electrical contact with the electrostatic gates. In this configuration, electron–hole pairs may still be generated in the vicinity of the resonator during illumination, but they rapidly recombine in the absence of a confining heterointerface~\cite{leo1987}. As a result, the region beneath the resonator remains effectively free of long-lived charge carriers, preserving its microwave performance.

The device is mounted on a printed circuit board and enclosed within a microwave cavity to suppress spurious electromagnetic modes. Measurements are performed in a dilution refrigerator at a base temperature of approximately 10~mK. Optical excitation is provided by a 785~nm laser coupled to an optical fiber routed into the dilution refrigerator. The fiber terminates in a collimator that directs the incoming light toward a $\sim$1~$\mu$m-wide pinhole in the copper enclosure surrounding the device. The resulting illumination spot is significantly larger than the mesa, ensuring uniform excitation of the entire heterostructure. All data presented in this work are acquired under these conditions using the setup illustrated in Fig.~\ref{fig:fig1}(a).

\section{\label{sec:illumination}Photogeneration of charge carriers}

The double quantum dot device is first characterized at base temperature, prior to optical illumination. The resonator frequency is identified by measuring the transmission parameter $S_{21}$ using a vector network analyzer (VNA). A fixed probe frequency is then selected near the resonance, where the transmission response is sensitive to changes in the resonance frequency. The transmission is subsequently recorded at this frequency while sweeping the gate voltages to obtain a charge stability diagram. When measured over a wide range of gate voltages, the diagram exhibits no discernible charge transition lines. This absence of features is consistent with an initially empty device, indicating that no free carriers are present at the heterointerface. In this regime, residual defects are either ionized and contribute to the Schottky potential, or remain frozen out at low temperature.

The next step is to generate electron–hole pairs in the GaAs layer using photons with energy exceeding the GaAs band gap (1.519~eV, corresponding to 816~nm), motivating the choice of a low-power laser with a wavelength of 785~nm. Since this photon energy remains below the band gap of the AlGaAs barrier, the generation of carriers in the barrier layer is minimized. During an illumination cycle, the laser is turned on for a fixed duration of 10~s. Importantly, this illumination is realized while keeping all gates (later referred as $V_\text{all}$) at 0~V. Based on independent power measurements, the photon flux at the device is estimated to be on the order of $10^{12}$~photons/s. A fraction of these photons is absorbed in the mesa region, resulting in the generation of electron–hole pairs.

\begin{figure*}[t]
    \centering
    \includegraphics[width=1\textwidth]{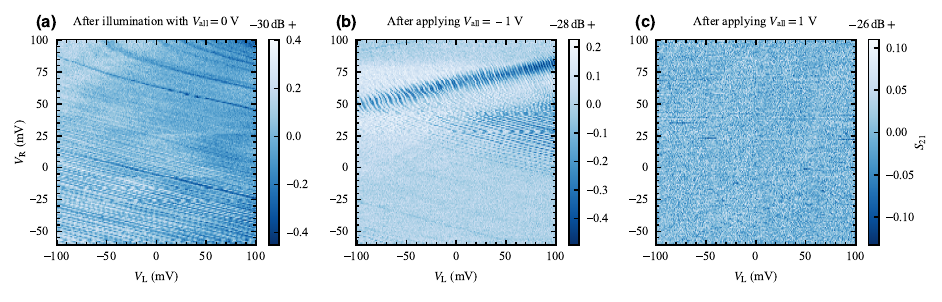}
    \caption{Charge stability diagrams after different gate biases. (a) Charge stability diagram measured after illumination with all gates held at $0~\mathrm{V}$. The middle gate is set to 0~V during the measurement. The diagonal lines correspond to charge transitions. (b) Same stability diagram after applying $-1~\mathrm{V}$ to all three gates for $30~\mathrm{s}$. Charge transitions remain visible. (c) Same as (b), but after applying $+1~\mathrm{V}$ to all three gates for $30~\mathrm{s}$. All charge transitions disappear, leaving an empty stability diagram.}

    \label{fig:fig2}
\end{figure*}

Figure~\ref{fig:fig2}(a) shows a stability diagram measured over the same large voltage range, but following an illumination cycle. In contrast to the pre-illumination measurement, multiple charge transition lines are now observed. At a given fixed-readout frequency, these features appear as a reduction of the resonator transmission and correspond to changes in the charge occupation of the gate-defined potential landscape. The uniform slope of the transition lines indicates that the electrostatic confinement is dominated by a single effective potential, resulting from the combined influence of all three control gates. In this regime, the device behaves as a single, large quantum dot rather than two well-separated quantum dots. Achieving independent control of the two dots requires decreasing the coupling, which is obtained by increasing the barrier height via the middle gate. Before doing so, however, it is necessary to determine whether the trapped charge carriers are electrons or holes since the illumination took place while $V_\text{all}=0$~V.

The photogenerated charge carriers are confined at the heterointerface by the built-in potential associated with the Schottky contact. To determine their polarity, we exploit the fact that applying a sufficiently large gate voltage can invert the band curvature at the interface. Such an inversion is expected to expel the confined carriers into the bulk of the substrate, where they are no longer trapped and do not return upon restoring the gate voltage to 0~V, thereby effectively resetting the device.

Experimentally, applying a negative voltage of $-1$~V to all gates results in only minor modifications to the electrostatic landscape, and charge transition lines remain clearly visible in Fig.~\ref{fig:fig2}(b). In contrast, applying a positive voltage of $+1$~V leads to the complete disappearance of all transition lines over the full voltage range, as shown in Fig.~\ref{fig:fig2}(c). This observation indicates that only a positive bias can invert the band curvature and effectively suppress the trapping region at the heterointerface necessary to maintain the charge carriers. These results strongly suggest that the confined carriers are holes.

\begin{figure}[ht]
    \centering
    \includegraphics[width=246pt]{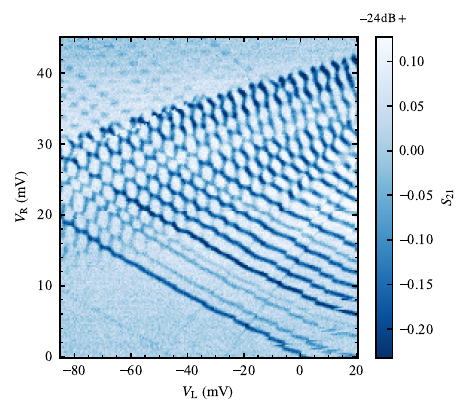}
    \caption{Charge stability diagram of the double quantum dot. Following an initial illumination with $V_\text{all}$, the left and right plunger gates are swept while the middle gate is held at $1~\mathrm{V}$. This increases the tunnel barrier height, allowing the characteristic hexagonal cells of a double quantum dot to be resolved.}

    \label{fig:fig3}
\end{figure}

Based on the identification of hole carriers, a positive bias can be applied to the middle gate to reduce the tunnel coupling between the two dots while simultaneously depleting the leads beneath the plunger gates. The experimental sequence is as follows: the device is first illuminated with all gates set to 0~V, after which a positive voltage of 1~V is applied to the middle gate. A charge stability diagram measured under these conditions is shown in Fig.~\ref{fig:fig3}. In contrast to the previously observed single-dot behavior, a clear hexagonal charge stability pattern emerges, indicating the formation of a well-defined double quantum dot with reduced interdot coupling~\cite{vanderwiel2002}. Similar behavior is observed for middle gate voltages ranging from 0.8~V to 1.2~V, with only minor shifts in the overall electrostatic landscape, as expected given the strong capacitive coupling of the middle gate to the double dot potential.

The hexagonal cells are relatively small compared to the linewidth of the charge transitions. This is consistent with the large size of the plunger gates, which leads to a significant total capacitance $C_\text{tot}$ and, consequently, a reduced charging energy $E_\text{C} = e^2 / C_\text{tot}$. Moreover, the single-particle level spacing decreases as the quantum dot size increases. Consequently, larger dots exhibit smaller addition energies and more closely spaced charge transitions~\cite{vanderwiel2002}.

Another notable feature is the absence of some transition lines in specific regions of the charge stability diagram. These missing transitions are most likely due to a reduced sensitivity of the coupled coplanar waveguide resonator to slow charge dynamics. Similar behavior has been reported in previous studies and attributed to low tunneling rates~\cite{frey2012b, frey2012,sokolov2025}. When the tunnel rate is much smaller than the resonator frequency, the charge cannot follow the RF excitation, strongly suppressing the dispersive response~\cite{demichielis2023}. As a result, the three types of transitions (two intradots and one interdot) can exhibit different effective visibility, depending on their respective tunneling rates. These rates depend on the gate voltages, hence the different visibility regions.

\section{\label{sec:schottky}Role of the Schottky contact}

Even without any applied gate voltages, our device presents a trapping region at the heterointerface. Although not initially intended, this built-in potential at the metal–semiconductor interface arises from the formation of a deep Schottky contact in an otherwise undoped semiconductor, aside from residual doping associated with native defects~\cite{hurle1999,wenzl1993,martin1977,mitonneau1977}. Assuming a uniform distribution of ionized donor-like defects, the electrostatic potential inside the semiconductor can be obtained by solving Poisson’s equation. This yields a parabolic band bending, from which the conduction band edge can be expressed as

\begin{equation}
E_\text{c}(x) = V_\text{b} \left[1 - \left(\frac{x}{d}\right)\right]^2,
\end{equation}

where $x$ is the depth along the growth direction, $V_\text{b}$ is the Schottky barrier energy, and $d$ is the characteristic depletion length given by

\begin{equation}
d = \sqrt{\frac{2 \epsilon_0 \epsilon_\text{r} V_\text{b}}{e^2 N_\text{D}}},
\end{equation}

where $e$ is the elementary charge, $\epsilon_0$ the vacuum permittivity, $\epsilon_\text{r}$ the relative permittivity of the semiconductor, and $N_\text{D}$ the donor density~\cite{davies1998}. The relatively low density of native defects results in a large depletion length~\cite{dubecky1994}, allowing the band bending to extend to the AlGaAs/GaAs interface. Locating the trapping potential within the depletion region minimizes the equilibrium density of mobile charge carriers, providing an initially empty environment in which photogenerated carriers can be trapped and used to populate the quantum dot.

To obtain a more realistic description of the electrostatic landscape, we performed finite-element simulations using the Quantum Technology Computer-Aided Design (QTCAD) software developed by Nanoacademic Technologies. This framework can solve the nonlinear Poisson equation in complex geometries, providing insight into how the combined effect of the Schottky potential and gate voltages shapes the confinement potential of the quantum dots. The simulation requires a three-dimensional mesh of the heterostructure, onto which appropriate boundary conditions are applied, with the metallic gates treated as Schottky contacts. This boundary condition is given by~\cite{nanoacademic}

\begin{equation}
    \phi = e\Delta V - \frac{1}{e}\left(V_\text{b}+\chi\right),
\end{equation}

where $\phi$ is the potential at the boundary, $\Delta V$ is the voltage applied to the metallic gate, $V_\text{b}$ is the Schottky barrier energy, fixed at $0.7$~eV and $\chi$ is the GaAs electron affinity, fixed at $4.07$~eV.

\begin{figure}[ht]
    \centering
    \includegraphics[width=246pt]{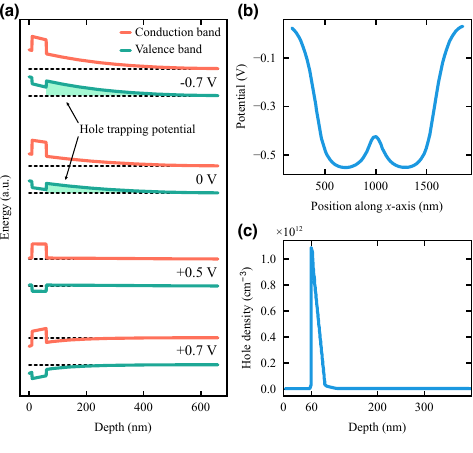}
    \caption{Results of the finite-element simulation. (a) Band profiles along the $z$-axis for different gate voltages. The black dashed line is a visual guide to highlight the curvature of the bands. (b) Line cut of the potential along the $x$-axis at the AlGaAs/GaAs heterojunction beneath the two plunger gates, with $V_\text{L} = V_\text{R} = -0.5$~V and $V_\text{M} = 0.5$~V. (c) Hole density along the $z$-axis for the same gate voltages as in (b). The bottom of the device is defined as an ohmic boundary, providing the carrier reservoir required to reproduce the hole population expected following illumination.}

    \label{fig:fig4}
\end{figure}

Figure~\ref{fig:fig4}(a) illustrates the influence of the gate voltages on the band profile. For reference, typical intentional doping densities in similar heterostructures are ranging from $10^{16}$~cm$^{-3}$~\cite{morley1992} to $10^{18}$~cm$^{-3}$\cite{babin2021}, whereas the value used here, $N_\text{D} = 5 \times 10^{14}$~cm$^{-3}$, is significantly lower and representative of residual defect densities in molecular beam epitaxy (MBE)–grown materials ~\cite{larkins1986}. At zero applied gate voltage, the intrinsic band bending produces a confinement potential for holes, assuming predominantly donor-like defects which is often the case for low-carbon, undoped MBE growth. Applying a negative voltage (e.g., $-0.7$~V) enhances this confinement, while a positive voltage (e.g., $+0.7$~V) progressively reduces and eventually inverts the band curvature, thereby suppressing the confining potential. This behavior is consistent with the experimental observation that applying a sufficiently large positive bias (1~V) empties the quantum dots. The simulations further indicate that the flat-band condition occurs at approximately 0.5~V, corresponding to the transition between accumulation and depletion regimes.

Figure~\ref{fig:fig4}(b) illustrates a line cut of the potential at the heterojunction beneath the two plunger gates. For this simulation, the left and right gates were set to $-5$~V, while the middle gate was set to $+5$~V. This extreme biasing condition is used to demonstrate that substantial decoupling of the two quantum dots can be achieved with this gate geometry, allowing holes to be confined independently in each dot. Figure~\ref{fig:fig4}(c) shows the hole density beneath one of the plunger gates as a function of depth. A sharp increase in the hole density at 60~nm below the surface indicates the formation of a hole gas. In this simulation, the bottom face of the 3D mesh was assigned an ohmic boundary condition, providing the carrier reservoir required to reproduce the hole population expected following illumination.

\section{\label{sec:conclu}Conclusion}

This work demonstrates the ability to populate gate-defined quantum dots using optically generated charge carriers. This approach eliminates the need for ohmic contacts, in-situ doping, and engineered charge reservoirs, relying instead on the generation of electron-hole pairs within the semiconductor. While the present work does not yet demonstrate control of a spin qubit, it establishes the charge control required as a foundation for such devices. The next important step is to reach the single-charge regime. A straightforward path toward this goal is to optimize the gate geometry by reducing the size of the plunger gates and employing a narrower gate to control the interdot tunnel coupling. Smaller quantum dots combined with a sharper tunneling barrier should provide the additional confinement and control required to operate the double quantum dot in the single-charge regime. Further improvements could be achieved by optimizing the optical setup to reduce the number of photons reaching the device and to more precisely control the illuminated region. Such improvements would provide finer control over the number and spatial distribution of the photogenerated charges, bringing the device closer to deterministic charge initialization.

Beyond charge initialization, the ability to optically populate a quantum dot opens new possibilities for interfacing semiconductor qubits with photonic and microwave systems. In particular, it could enable the transduction of a quantum state encoded in an incoming photon into the spin state of a trapped charge. This state could subsequently be transferred to a microwave photon in a coupled resonator, providing a potential interface between semiconductor spin qubits and superconducting quantum systems. Moreover, photo-population is naturally compatible with flip-chip architectures, since it removes the need for ohmic contacts on the heterostructure chip. This could allow the resonator and its associated fabrication steps to be implemented entirely on a separate, high-$Q$ substrate, such as silicon or sapphire, while leaving the semiconductor heterostructure largely unaltered. Together, these features highlight the potential of optical charge initialization as a route toward scalable semiconductor quantum devices and hybrid quantum architectures.

\begin{acknowledgments}
This work was made possible through a collaboration with the National Research Council of Canada (NRC) under the Quantum Sensors (QSP) Challenge Program and the Quantum Internetworking (QUIN) Challenge program. Part of this research was also funded by the Fonds de recherche du Québec (FRQ), The Canada First Research Excellence Fund (CFREF) and Photonique Quantique Québec (PQ2). We would like to acknowledge the support of our collaborators at the NRC, in particular Pedro Barrios and Jean Lapointe for their expertise in microfabrication, and Marek Korkusinski for valuable discussions. We also acknowledge the support of researchers at Nanoacademic Technologies, in particular Félix Beaudoin and Pericles Philippopoulos. Finally, we thank all the students and interns who contributed to this research project.
\end{acknowledgments}




\bibliography{bibliography}

\end{document}